\documentclass[10pt,preprint]{aastex}

\newcommand{\hd}{HD~209458}

\def\ltsima{$\; \buildrel < \over \sim \;$}
\def\lsim{\lower.5ex\hbox{\ltsima}}
\def\gtsima{$\; \buildrel > \over \sim \;$}
\def\gsim{\lower.5ex\hbox{\gtsima}}

\slugcomment{Accepted for publication in the Astrophysical Journal}
\shorttitle{Spin-Orbit Alignment in \hd}
\shortauthors{Winn et al.}

\begin{document}

\title{ Measurement of Spin--Orbit Alignment in\\
an Extrasolar Planetary System}

\author{
Joshua N.\ Winn\altaffilmark{1,2},
Robert W.\ Noyes\altaffilmark{1},
Matthew J.\ Holman\altaffilmark{1},
David Charbonneau\altaffilmark{1},\\
Yasuhiro Ohta\altaffilmark{3},
Atsushi Taruya\altaffilmark{3,4},
Yasushi Suto\altaffilmark{3,4},
Norio Narita\altaffilmark{3},
Edwin L.\ Turner\altaffilmark{3,5},\\
John A.\ Johnson\altaffilmark{6},
Geoffrey W.\ Marcy\altaffilmark{6},
R.\ Paul Butler\altaffilmark{7},
Steven S.\ Vogt\altaffilmark{8}
}

\altaffiltext{1}{Harvard-Smithsonian Center for Astrophysics, 60
Garden Street, Mail Stop 51, Cambridge, MA 02138}

\altaffiltext{2}{Hubble Fellow}

\altaffiltext{3}{Department of Physics, The University of Tokyo,
Tokyo, 113-0033, Japan}

\altaffiltext{4}{Research Center for the Early Universe, School of
Science, The University of Tokyo, Tokyo 113-0033, Japan}

\altaffiltext{5}{Princeton University Observatory, Peyton Hall,
Princeton, NJ 08544}

\altaffiltext{6}{Department of Astronomy, University of California,
Mail Code 3411, Berkeley, CA 94720}

\altaffiltext{7}{Department of Terrestrial Magnetism, Carnegie
Institution of Washington, 5241 Broad Branch Road NW, Washington, DC
20015}

\altaffiltext{8}{UCO/Lick Observatory, University of California, Santa
Cruz, CA 95064}

\begin{abstract}
We determine the stellar, planetary, and orbital properties of the
transiting planetary system HD~209458, through a joint analysis of
high-precision radial velocities, photometry, and timing of the
secondary eclipse. Of primary interest is the strong detection of the
Rossiter-McLaughlin effect, the alteration of photospheric line
profiles that occurs because the planet occults part of the rotating
surface of the star. We develop a new technique for modeling this
effect, and use it to determine the inclination of the planetary orbit
relative to the apparent stellar equator ($\lambda = -4\fdg4 \pm
1\fdg4$), and the line-of-sight rotation speed of the star ($v\sin
I_\star = 4.70\pm 0.16$~km~s$^{-1}$). The uncertainty in these
quantities has been reduced by an order of magnitude relative to the
pioneering measurements by Queloz and collaborators. The small but
nonzero misalignment is probably a relic of the planet formation
epoch, because the expected timescale for tidal coplanarization is
larger than the age of the star. Our determination of $v\sin I_\star$
is a rare case in which rotational line broadening has been isolated
from other broadening mechanisms.
\end{abstract}


\section{Introduction}

A star and its planets inherit their angular momentum from a common
source:\ the rotation of the molecular cloud from which they
formed. It follows that the axes of planetary orbits should be closely
aligned with the rotation axis of the star, an expectation that is
fulfilled in the solar system. The rotation axis of the Sun is tilted
by only 6\arcdeg\ relative to the axis defined by the net angular
momentum of the planetary orbits (see Beck \& Giles 2005, and
references therein). Indeed, the observed coplanarity of solar system
orbits was the main inspiration for Kant (1755) and Laplace (1796),
who proposed that the Sun and its planets condensed from a spinning,
flattened nebula.

It would be interesting to know whether this degree of alignment is
typical of all planetary systems, or whether the solar system is
anomalously well-aligned or misaligned. The degree of alignment
depends on both poorly understood initial conditions and poorly
understood physical processes. For example, the angular momentum
distribution of the parent molecular cloud is surely inhomogeneous at
some level. The star forms earlier than the planets, and might
consequently be composed of material with a different net angular
momentum than the material that falls in later. The star's axis of
rotation may be altered during the T~Tauri phase, when much of its
angular momentum is lost through bipolar outflows and magnetic
coupling to the protoplanetary disk. There may be angular momentum
evolution in the orbits during planetary migration, or as a
consequence of gravitational interactions between protoplanets. The
orbits may be altered by tidal interactions with the parent star, or
torques from a companion star. It is even conceivable that there are
planetary systems for which the orbital axes are grossly misaligned
with the stellar rotation axis, due to a close encounter with another
star, or the outright capture of planets from another star. The
discovery of even a single example of such a system would be of
interest. Thus, it would be desirable to have additional cases besides
the solar system for which the degree of spin-orbit alignment can be
assessed.

Furthermore, if it could be established that planetary orbits are
universally well-aligned with the equatorial planes of their parent
stars, there would be useful corollaries for some planet detection and
characterization schemes, as reviewed by Hale (1994). For instance,
the Doppler method does not reveal the mass $M_p$ of the planet,
but rather $M_p \sin I$, where $I$ is the inclination of the
orbit relative to the sky plane. It would be helpful if the
inclination of the stellar rotation axis $I_\star$ could be safely
assumed to equal $I$, because then the various methods of estimating
$I_\star$ (through estimates of rotation periods, line broadening, and
stellar radius) could be brought to bear on the problem. Indeed, some
investigators have already found it convenient to assume $I=I_\star$
in interpreting radial velocity data (e.g., Hale 1995, Francois et
al.\ 1996, Baliunas et al.\ 1997, Gonzalez 1998). Likewise, if
$I=I_\star$, then stars that are viewed pole-on ($I_\star\approx
0\arcdeg$) would be good targets for direct-detection experiments,
since the planets would always be viewed near maximum elongation.

For transiting planets, there is a powerful method available to
measure spin-orbit alignment. The idea is to exploit a spectroscopic
effect that was first described by Rossiter (1924) and McLaughlin
(1924) for the case of eclipsing binary stars. During a transit, the
planet occults a spot on the rotating surface of the star. The
occultation thereby removes a particular velocity component from the
rotationally broadened profiles of the stellar absorption lines. In
principle, through observation of this missing velocity component, one
can measure the line-of-sight velocity of the stellar disk ($v\sin
I_\star$) at each point along the chord traversed by the planet, as
well as the angle ($\lambda$) between the sky-projected angular
momentum vectors of the planetary orbit and the stellar spin.

In practice, the spectral distortion produced by the
Rossiter-McLaughlin (RM) effect is difficult to discern in a single
spectral line. However, when the entire spectrum is analyzed, the RM
effect manifests itself as an anomalous radial velocity. By
``anomalous,'' we mean an apparent wavelength shift of the spectral
lines that differs from the Doppler shift caused by the star's orbital
motion. If the planet blocks a small portion of the blue wing of the
line, then the line will appear to be slightly redshifted, and vice
versa. Calculations of this effect were carried out by Hosokawa
(1953), Kopal (1980), and most recently by Ohta, Taruya, \& Suto
(2005; OTS hereafter). The latter authors derived analytic expressions
for the case of planetary transits, and suggested that $\lambda$ could
be determined within a few degrees if high-precision
($\sim$5~m~s$^{-1}$) radial velocity measurements were obtained
throughout a transit.

Such high precision can only be achieved for bright stars, and the
brightest star that is known to host a transiting planet is HD~209458
(Henry et al.\ 2000, Charbonneau et al.\ 2000). This remains the only
exoplanetary system for which the RM effect has been detected. Three
different groups have reported detections: Queloz et al.\ (2000),
Bundy \& Marcy (2000), and Snellen (2004). Most pertinent to this
paper is the work of Queloz et al.\ (2000), who measured the apparent
radial velocity of the star during a planetary transit with a
precision of 10~m~s$^{-1}$, finding $\lambda$ to be consistent with
zero within about 20\arcdeg, and $v\sin I_\star=3.75\pm
1.25$~km~s$^{-1}$ along the transit chord.\footnote{Bundy \& Marcy
(2000) detected the effect but did not have enough transit data to
justify a thorough analysis. Snellen (2004) had a different
motivation. He assumed $\lambda=0\arcdeg$ and attempted to detect
absorption lines in the atmosphere of the planet through the
wavelength dependence of the RM effect.} Since this pioneering work, a
treasure-trove of new data has become available, including transit
photometry with $10^{-4}$ precision (Brown et al.\ 2001), radial
velocity measurements with 3--4~m~s$^{-1}$ precision (Laughlin et al.\
2005), and a recent measurement of the time and duration of the
secondary eclipse (Deming et al.\ 2005).

The motivation for the work described in this paper was to investigate
the degree to which the RM analysis could be improved using a
combination of these high-precision data. We have also taken the
opportunity to update the determinations of the other stellar,
planetary, and orbital parameters, on the basis of the joint analysis
of radial velocity measurements, photometry, and the timing of the
secondary eclipse.

The data on which our analysis is based are described in the following
section. The model that was used to fit the data is described in \S~3,
including a new and empirical method to calculate the anomalous radial
velocity due to the RM effect. In \S~4 we present the results, showing
in particular that the uncertainties in $\lambda$ and $v\sin I_\star$
have been reduced by an order of magnitude. We also remark on the
determination of the orbital eccentricity, a parameter that has been
of particular interest ever since Bodenheimer, Lin, \& Mardling (2001)
pointed out that ongoing eccentricity damping could produce enough
tidal heating to account for the unexpectedly large size of the
planet. Finally, in \S~5 we place the RM results in the context of
theories of tidal interactions between planets and their parent stars,
and consider the possible significance of this unusually direct means
of measuring the projected rotation speed of the star.

\section{The Data}

Our work is based on three types of data: (1) radial velocity
measurements of the parent star throughout its entire orbit, including
the transit phase; (2) optical photometry of the system during the
transit phase; (3) infrared photometry of the system during secondary
eclipse (the phase when the planet is behind the star). We have not
obtained any new data. Instead, we have chosen the highest-precision
data that are currently available in each of these categories.

\subsection{Radial velocities}
\label{subsec:data_rv}

Our radial velocity measurements are from Laughlin et al.\ (2005), who
used the Keck~I 10~m telescope and the High Resolution Echelle
Spectrograph (HIRES) equipped with an iodine cell for accurate
wavelength calibration. A total of 85 spectra were acquired between
November 1999 and December 2004. They were generally taken at random
orbital phases, with the notable exception of UT~2000~July~29, when a
sequence of 17 spectra was acquired during a transit. This
transit-phase subset is obviously important for this work. The rest of
the data is also important, because the Keplerian orbit must be known
with high accuracy in order for the radial velocity anomaly to be
isolated and interpreted. We did not use the three spectra that were
taken during ingress or egress because our model for the RM effect,
which is presented in \S~\ref{subsec:rv}, is least accurate during the
partial phases of the transit.

The radial velocities were derived from the spectra using the
technique described by Butler et al.\ (1996), and have a typical
measurement error of 3--4~m~s$^{-1}$. The zero point of the radial
velocity scale is arbitrary. As noted by Laughlin et al.\ (2005), a
star such as HD~209458 should produce intrinsic radial velocity noise
with a standard deviation of approximately 2.8~m~s$^{-1}$, an
empirical estimate based on radial-velocity and chromospheric
monitoring of similar stars (Saar, Butler, \& Marcy 1998). This
intrinsic noise, often referred to as ``stellar jitter,'' presumably
arises from motions or flux variations of the stellar surface. For
this reason, as an estimate of the total uncertainty in each radial
velocity, we added 2.8~m~s$^{-1}$ in quadrature to the quoted
measurement error.

\subsection{Transit photometry}

Our photometry is from Brown et al.\ (2001), who used the {\it Hubble
Space Telescope} and the Space Telescope Imaging Spectrograph (STIS),
sadly now defunct, to record the flux of HD~209458 within a
$\approx$50~nm band centered on 610~nm. They achieved the
extraordinary precision of $10^{-4}$ in relative flux by using STIS as
a dispersive photometer. The resulting photometric time series is
divided into 20 segments: on each of four occasions (``visits''), the
star was observed for five orbits of the telescope around the
Earth. The visits were chosen to span particular transit events in
2000 April and May.

The first two orbits and the last orbit of each visit took place when
the planet was not transiting. Those data served only to establish the
flux baseline for the time variations observed in the third and fourth
orbits of each visit. Following Brown et al.\ (2001), we excluded from
consideration the data from the first orbit of each visit, reasoning
that they are unnecessary and perhaps even undesirable because the
telescope and instrument need time to settle into maximum
stability. We also excluded all the data from the first visit, because
those data were affected by an instrumental problem (see Brown et al.\
2001, \S~2). The resulting data set consists of 417 measurements of
relative flux, including excellent coverage of the entire transit
phase.

\subsection{Secondary eclipse timing}
\label{subsec:secondary}

The only successful detection of the secondary eclipse was recently
achieved by Deming et al.\ (2005), who monitored the 24~$\mu$m flux of
the system with the Multiband and Imaging Photometer aboard the {\it
Spitzer Space Telescope}. They detected the diminution of total flux
when the planet was hidden by the star, with a total signal-to-noise
ratio of 5--6. For our purpose, direct modeling of the light curve is
not very useful because of the low signal-to-noise ratio of each data
point. However, measurements of the time and duration of the secondary
eclipse are potentially useful in determining the orbital eccentricity
($e$) and argument of pericenter ($\omega$). The following expressions
are valid to first order in $e$:
\begin{eqnarray}
e\cos\omega & = & \frac{\pi}{2P} \left( t_{\rm II} - t_{\rm I} - \frac{P}{2} \right) \\
e\sin\omega & = & \frac{\Theta_{\rm I} - \Theta_{\rm II}}{\Theta_{\rm I} + \Theta_{\rm II}},
\label{eq:secondary}
\end{eqnarray}
where $t_{\rm I}$ and $t_{\rm II}$ are the midpoints of the primary
eclipse (transit) and secondary eclipse, respectively, and
$\Theta_{\rm I}$ and $\Theta_{\rm II}$ are the corresponding
durations. The orbital period is $P$. Using $t_{\rm II}$ measured by
Deming et al.\ (2005) and $t_{\rm I}$ measured by Brown et al.\
(2001), we find
\begin{equation}
e\cos\omega = \left( -0.6 \pm 3.7 \right) \times 10^{-3},
\label{eq:ecosw}
\end{equation}
which adds a little information beyond what can be learned from the
radial velocities and transit photometry alone (see \S~4). The
constraint on $e\sin\omega$ is not useful, because of the relatively
large fractional uncertainty in $\Theta_{\rm II}$. We estimated
$\Theta_{\rm II}$ from Fig.~1b of Deming et al.\ (2005) and found
\begin{equation}
e\sin\omega = -0.02 \pm 0.19,
\end{equation}
which is not sufficiently precise to improve on the constraints from
the spectroscopic orbit (Mazeh et al.\ 2000, Laughlin et al.\ 2005).

\section{Description of the Model}

\subsection{The Orbit}
\label{subsec:orbit}

\begin{figure}[h]
\epsscale{0.75}
\plotone{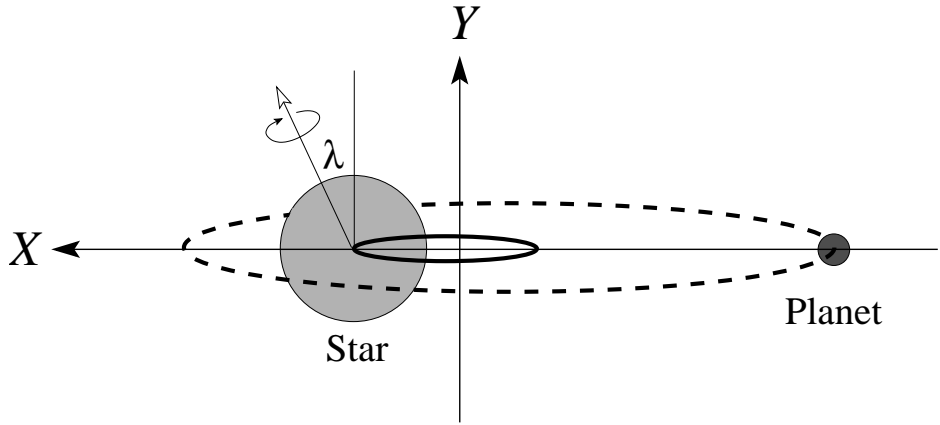}
\caption{ Coordinate system of the model. The $X$--$Y$ plane is the
sky plane. The $Z$ axis points away from the observer. The projected
stellar orbit is shown by the solid ellipse, and the projected
planetary orbit is shown by the dashed ellipse. Without loss of
generality, the longitude of nodes is chosen to lie along the $X$
axis. The argument of pericenter is zero in this illustration, but it
is a free parameter in the model. The angle $\lambda$ is between the
$Y$ axis and the sky projection of the stellar rotation axis.
\label{fig:diagram}}
\end{figure}

The basis of the model is a two-body Keplerian orbit (see
Fig.~\ref{fig:diagram}). The orbit is specified by the masses of the
star and planet ($M_{\star}$ and $M_p$), the orbital period
($P$), the orbital eccentricity ($e$), the argument of pericenter
($\omega$), the orbital inclination ($I$), and the radial velocity of
the center of mass ($\gamma$).\footnote{Laughlin et al.\ (2005)
reported only the radial velocity variations relative to an arbitrary
velocity standard. Our model parameter $\gamma$ is the offset between
this zero point and the heliocentric radial velocity of the center of
mass.} Given the initial condition, it is a venerable and
straightforward problem to compute the sky position $(X,Y)$ and radial
velocity $\dot{Z}$ of the center of mass of either body at any
subsequent time. We chose to parameterize the initial condition by the
free parameter $\Delta t_{\rm I}$, defined as
\begin{equation}
\Delta t_{\rm I} \equiv t_{\rm I} - {\rm 2,451,659.93675},
\end{equation}
where $t_{\rm I}$ is the central transit time, measured in
heliocentric Julian days. The reference time is the central transit
time measured by Brown et al.\ (2001). We did not allow $P$ to vary,
since it has been determined independently with much greater precision
than can be achieved with only the data analyzed here. We adopted the
value $P=3.52474895$~days, which is based on an analysis of the STIS
data by Brown et al.\ (2001) and a subsequent series of STIS
observations by Charbonneau et al.\ (2003). The uncertainty in $P$ is
83~ms which is negligible for our purposes (see \S~\ref{subsec:rm}).

\subsection{The Flux}
\label{subsec:flux}

There are two more parameters for the planetary radius ($R_p$) and
stellar radius ($R_\star$). The transit occurs when the sky position
of the planet is less than $R_p + R_{\star}$ away from the sky
position of the star, and $Z_\star > Z_p$. Outside transit, the model
flux is unity. During transit, the model flux is computed under the
assumption of a quadratic limb-darkening law,
\begin{equation}
\frac{I(\mu)}{I(1)} = 1 - u_1(1-\mu) - u_2(1-\mu)^2,
\label{eq:limbdark}
\end{equation}
using the algorithm of Mandel \& Agol (2002). In this expression,
$\mu$ is the cosine of the angle between the line of sight and the
normal to the stellar surface. The two numbers $u_1$ and $u_2$ are
free parameters. This is the same parameterization that was used by
Brown et al.\ (2001).

\subsection{The Radial Velocity}
\label{subsec:rv}

Outside transit, the model radial velocity of the star is
$\gamma+\dot{Z}$. During transit, we must take into account the RM
effect. We write the model radial velocity as $\gamma+\dot{Z}+\Delta
v$, where $\Delta v$ is the anomalous Doppler shift. The value of
$\Delta v$ obviously depends on the position of the planet and the
rotation rate of the star, but what is the exact dependence? Since
this is the heart of the matter, we devote extra attention to this
aspect of the model.

Queloz et al.\ (2000) interpreted their radial velocity measurements
using a finite-element model of the star. They divided the stellar
disk into 90,000 cells and calculated the expected surface brightness
and spectral line profile from each cell individually. Then, to
simulate a transit spectrum, they summed the spectra of the cells that
are not obstructed by the planet, weighted by the relative intensities
of the cells. Finally, to compute $\Delta v$, they measured the
apparent Doppler shift of this simulated spectrum, presumably using
the same reduction pipeline that they used on the actual data.

Ohta, Taruya, \& Suto (2005) derived an analytic formula for $\Delta
v$ as a function of the planet position and stellar rotation
rate. Using an analytic formula is much easier and computationally
faster than using a finite-element model. However, we were concerned
that the OTS formula is based on a premise that may not apply in this
case. The spectral distortion due to the RM effect is the subtraction
of a small fraction [$\propto (R_p/R_\star)^2$] of a narrow
range of velocities from the rotationally-broadened line profile. The
OTS formula gives the first moment of the distorted line profile (the
``center of gravity'' in wavelength space). Intuitively, one would
expect the first moment to be a good approximation to the measured
Doppler shift, but the Butler et al.\ (1996) method is not
specifically designed to measure the first moment of spectral
lines. Rather, it is designed to find the optimal value of an overall
wavelength shift that brings the observed spectrum and a template
spectrum into best agreement, while also fitting for many other free
parameters describing variations in the instrumental response. This
method is not the same as measuring a first moment; in particular, it
assumes that any spectral distortion is instrumental. When presented
with an intrinsically distorted transit spectrum, the degrees of
freedom that are intended to mimic instrumental changes may absorb
some of the true signal. For this reason, the applicability of the OTS
formula is not assured.

Our approach to this problem was to test the OTS formula in a manner
similar in spirit to the technique of Queloz et al.\ (2000). We
simulated Keck/HIRES spectra taken during a transit; then, we
``measured'' $\Delta v$ from these simulated spectra using the same
reduction pipeline that was used by Laughlin et al.\ (2005); and
finally, we compared the results with the predictions of the OTS
formula. The details were as follows. We began with the NSO solar
spectrum (Kurucz, Furlenid, \& Brault 1984), which for our purposes
has effectively infinite resolution, and performed the following
steps:
\begin{enumerate}

\item Broaden the NSO spectrum to mimic the disk-integrated spectrum
of HD~209458. The convolution kernel was chosen so that the broadened
spectrum had a line width of 4.5~km~s$^{-1}$, the value measured by
Fischer \& Valenti (2005) for HD~209458, and was computed assuming a
linear limb-darkening law appropriate for the star's color and the
mean wavelength of the relevant spectral region ($u_1=0.6$, $u_2=0$;
Gray 1992). Call this spectrum $S_\star$.

\item Begin again with the NSO spectrum at the native (unbroadened)
line width. Scale it by an overall factor $f$, Doppler-shift it by an
amount $\Delta\lambda/\lambda = v_p/c$, and refer to the result
as $S_p$. This is meant to represent the spectrum of the
occulted portion of the stellar disk: $f$ is the flux of the occulted
portion, and $v_p$ is the mean line-of-sight velocity (the
``sub-planet'' velocity) of the occulted portion.

\item Compute $S_{\rm tr} = S_\star - S_p$, where ``tr'' indicates
``transit.'' This is the simulated transit spectrum at infinite
resolution.

\item Multiply $S_{\rm tr}$ by the measured iodine absorption
spectrum, which is also effectively of infinite resolution.

\item Convolve the result with a model point-spread function that is
derived from actual Keck/HIRES observations of HD~209458, and store
the result in the same digital format as reduced Keck/HIRES spectra.

\item Use the result as input to the same Doppler-shift measuring
algorithm that was employed on the actual Keck/HIRES spectra. Record
the result as $\Delta v$.

\end{enumerate}

This procedure does not account for any time variations of $f$ and
$v_p$ during the spectroscopic exposure, but OTS have shown that
such time variations are negligible for exposure times less than 10
minutes. We performed the preceding steps for different choices of the
input parameters $f$ and $v_p$, producing a two-dimensional grid
of results $\Delta v(f,v_p)$. We allowed $f$ to vary from zero
to 0.02, and $v_p$ to vary from $-4.5$ to
$+4.5$~km~s$^{-1}$. The resulting surface is very well described by a
polynomial approximation,
\begin{equation}
\Delta v = -f v_p
\left[
1.33 - 0.483\left( \frac{v_p}{{\rm 4.5~km~s}^{-1}} \right)^2
\right],
\label{eq:jj}
\end{equation}
with differences smaller than 0.5~m~s$^{-1}$ between the grid values
and the polynomial approximation.

In order to compare the OTS formula with the results of our
simulations, we calculated $\Delta v$ at each step of a planetary
transit, using each of the two schemes. We adopted parameters for the
planet and the star that are similar to those of the HD~209458
system. At each moment during the transit, we used the OTS formula to
calculate $\Delta v$ as a function of the planet coordinates. We also
computed $f$ using the limb-darkening law of Eq.~(\ref{eq:limbdark}),
and $v_p$ assuming solid-body rotation for the star, which then
allowed us to calculate $\Delta v$ using Eq.~(\ref{eq:jj}). Figure~2
compares the results of the two different methods of determining
$\Delta v$, as a function of the $X$ coordinate of the planet. The OTS
formula is a reasonable approximation of the simulated results, but it
underpredicts the magnitude of $\Delta v$ by approximately 10\%.

\begin{figure}[h]
\epsscale{0.75}
\plotone{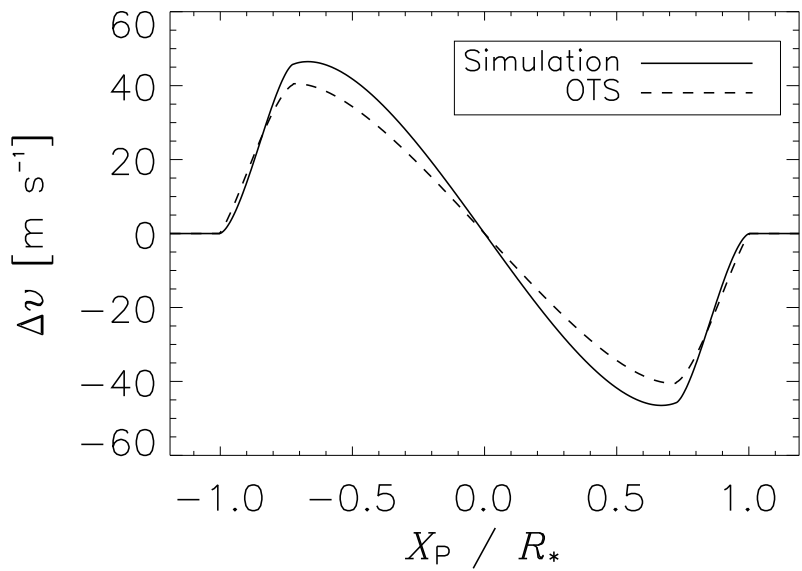}
\caption{ The radial velocity anomaly due to the Rossiter-McLaughlin
effect, as calculated {\it via} our simulations (solid line) and the
formula of Ohta et al.\ (2005) (dashed line). The horizontal axis is
the distance from the planet to the projected axis of the stellar
spin. The orbital, planetary, and stellar properties were chosen to be
approximately those of HD~209458: $R_p/R_\star = 0.12$, $Y_{\rm
P}/R_\star = 0.5$, $e=0$, $\lambda=0\arcdeg$, $v\sin
I_\star=4.5$~km~s$^{-1}$, and $u_1+u_2=0.64$.
\label{fig2}}
\end{figure}

We do not know the cause of the discrepancy, apart from the general
argument presented earlier that the quantity calculated by OTS is not
the quantity that is truly measured by the algorithm of Butler et al.\
(1996). In fact, we expected the OTS formula to {\it over}estimate the
magnitude of $\Delta v$, given our previous argument that some of the
free parameters in the Butler et al.\ (1996) algorithm might act to
dilute the signal. We believe that our simulations provide a better
representation of the true measurement, and hence, we employed
Eq.~(\ref{eq:jj}) in our model of the RM effect. However, as mentioned
in \S~\ref{subsec:data_rv}, we did not attempt to fit the three radial
velocity measurements that were taken during an ingress. This is
because during the partial transit phases, the planet is closest to
the stellar limb, which is when unmodeled physical effects are most
pronounced. Such effects include departures from quadratic
limb darkening, variations of the limb darkening across the area
covered by the planet, and any intrinsic center-to-limb variations of
the line profile, including the ``convective blue shift'' (Beckers \&
Nelson 1978).

In summary, the radial velocity anomaly of the model is calculated as
follows. At a given time during a transit, the positions of the star
and planet are determined, and $f$ is computed with the procedure
described above. The sub-planet velocity $v_p$ is computed by assuming
solid-body rotation of the star (an assumption whose validity is
considered briefly in \S~\ref{sec:discussion}). Then,
Eq.~(\ref{eq:jj}) is used to calculate $\Delta v$. The description of
the RM effect adds two additional free parameters to the model: $v\sin
I_\star$ of the stellar rotation, and $\lambda$, the angle between the
sky-projected axes of the planetary orbit and the stellar rotation
(see Fig.~\ref{fig:diagram}).

\section{Fitting Procedure and Results}
\label{sec:results}

The model has 13 free parameters, but only 12 of these parameters can
be determined independently. There is a well known degeneracy between
$M_\star$, $M_p$, and $R_\star$ as determined from transit data.
We chose to fix the value of $M_\star$ and optimize all the other
parameters. We repeated the optimization for three different choices
of $M_\star/M_\odot:$ $0.93$, $1.06$, and $1.19$. These values span
the full range of possibilities that Cody \& Sasselov (2002) concluded
is reasonable for HD~209458.

We used an AMOEBA algorithm (Press et al.\ 1992) to minimize
\begin{equation}
\chi^2 = \sum_{n=1}^{N_v} \left( \frac{v_O - v_C}{\sigma_v} \right)^2 +
         \sum_{n=1}^{N_f} \left( \frac{f_O - f_C}{\sigma_f} \right)^2 +
         \left( \frac{t_{{\rm II},O} - t_{{\rm II},C}}{\sigma_t} \right)^2
\end{equation}
as a function of all the parameters. Here, $v_O$ and $v_C$ are the
observed and calculated radial velocities, of which there are $N_v=83$
(14 during the transit phase). Likewise, $f_O$ and $f_C$ are the
observed and calculated fluxes, of which there are $N_f=417$. The
final term in the sum represents the constraint on the time of
secondary eclipse.

\begin{figure}
\epsscale{1.0}
\plotone{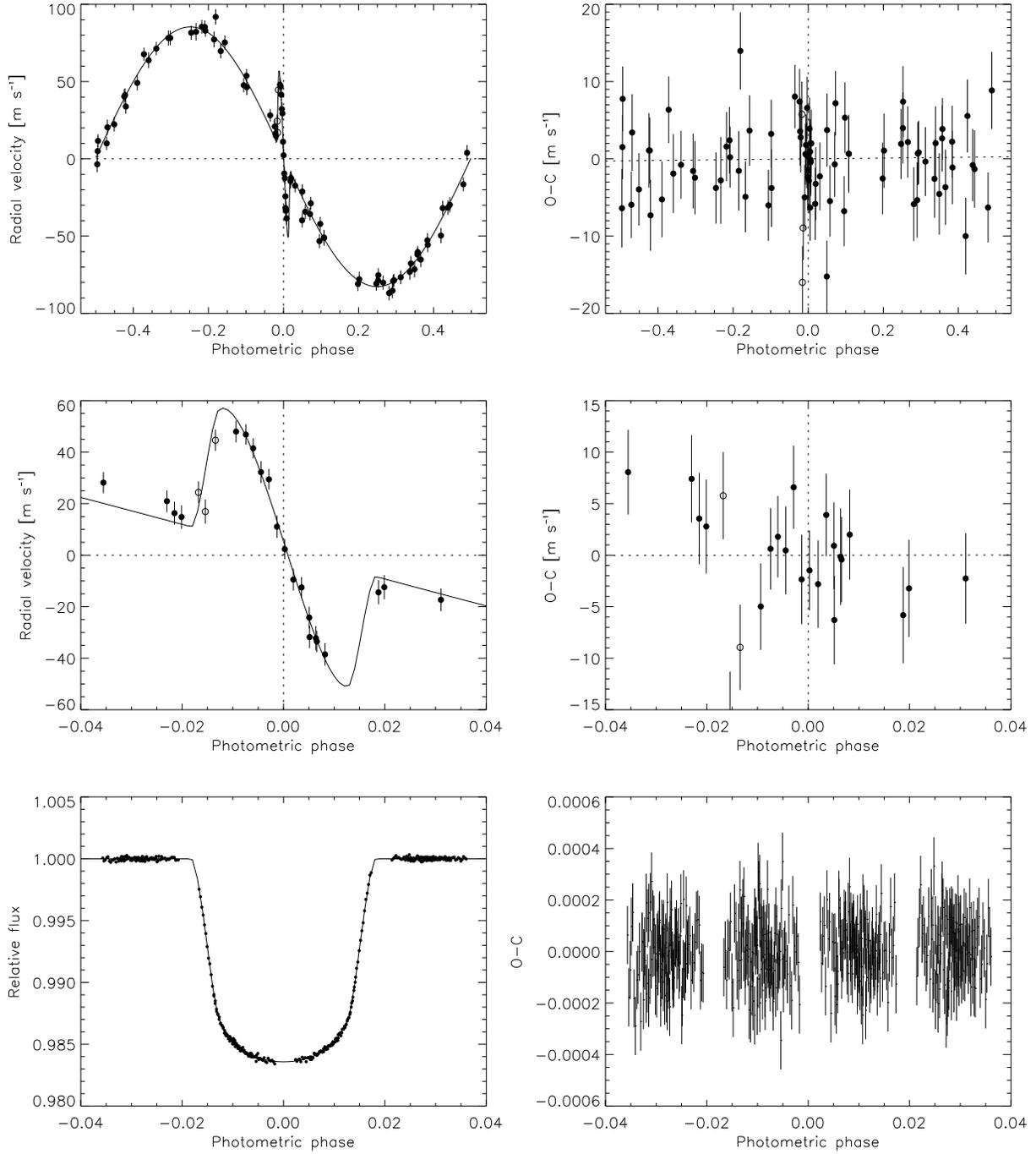}
\caption{ Comparison of the data and the best-fitting model, for the
case $M_\star/M_\odot=1.06$. The left panels show the data (points)
and the model (line), as a function of photometric phase. The right
panels show the residuals (observed minus calculated). {\it Top
row.}---The radial velocity data. {\it Middle row.}--- Close-up of the
radial velocity data near the transit phase, including 3 points (open
circles) that were not used in the fitting procedure. {\it Bottom
row.}---Photometry.
\label{fig:bestfit}}
\end{figure}

For $M_\star/M_\odot = 1.06$, the best-fitting model has $\chi^2 =
528$, with 489 degrees of freedom ($\chi^2/N_{\rm DOF} = 1.08$). We
consider this an excellent fit. The data and the best-fitting model
are compared in Fig.~\ref{fig:bestfit}. The RM effect is apparent as
the sinusoidal glitch in the radial velocity curve near zero
phase. The three radial velocity measurements that were taken during
ingress (and which were not used in the fitting procedure) are plotted
with open circles.

To estimate the uncertainties in the parameters, we performed a
bootstrap Monte Carlo analysis as described by Press et al.\
(1992). We created synthetic data sets, each of which had the same
$N_v$ and $N_f$ as the actual data set. Each entry in a synthetic data
set was a datum (a calendar date and the value of the flux or radial
velocity measured on that date) drawn randomly from the real data set,
with repetitions allowed. Thus, a substantial fraction of the entries
in each synthetic data set are duplicated at least once, and receive
greater weight in the $\chi^2$ sum. The idea is to estimate the
probability distribution of the data using the observed data values
themselves, rather than choosing models for the underlying physical
process and for the noise. For each of the three different choices of
$M_\star$, we created $10^5$ synthetic data sets, and re-optimized the
12 free parameters for each synthetic data set. The resulting
distribution of best-fitting parameters was taken to be the joint
probability distribution of the true parameter values.

Table~1 gives the mean value of each parameter, the standard
deviation, and the estimated 90\% confidence limits. In addition to
the model parameters, results are given for some related derived
quantities, such as $R_p/R_\star$ and $e\cos\omega$. Most of the
results in Table~1 were based on the histograms of all $3\times 10^5$
results, i.e., they incorporate the uncertainty in the stellar
mass. For those parameters whose uncertainties are dominated by the
uncertainty in stellar mass, results are also given for the particular
case $M_\star/M_\odot=1.06$, to show how much the uncertainty would be
reduced with perfect knowledge of the stellar mass.

\begin{figure}
\epsscale{1.0}
\plotone{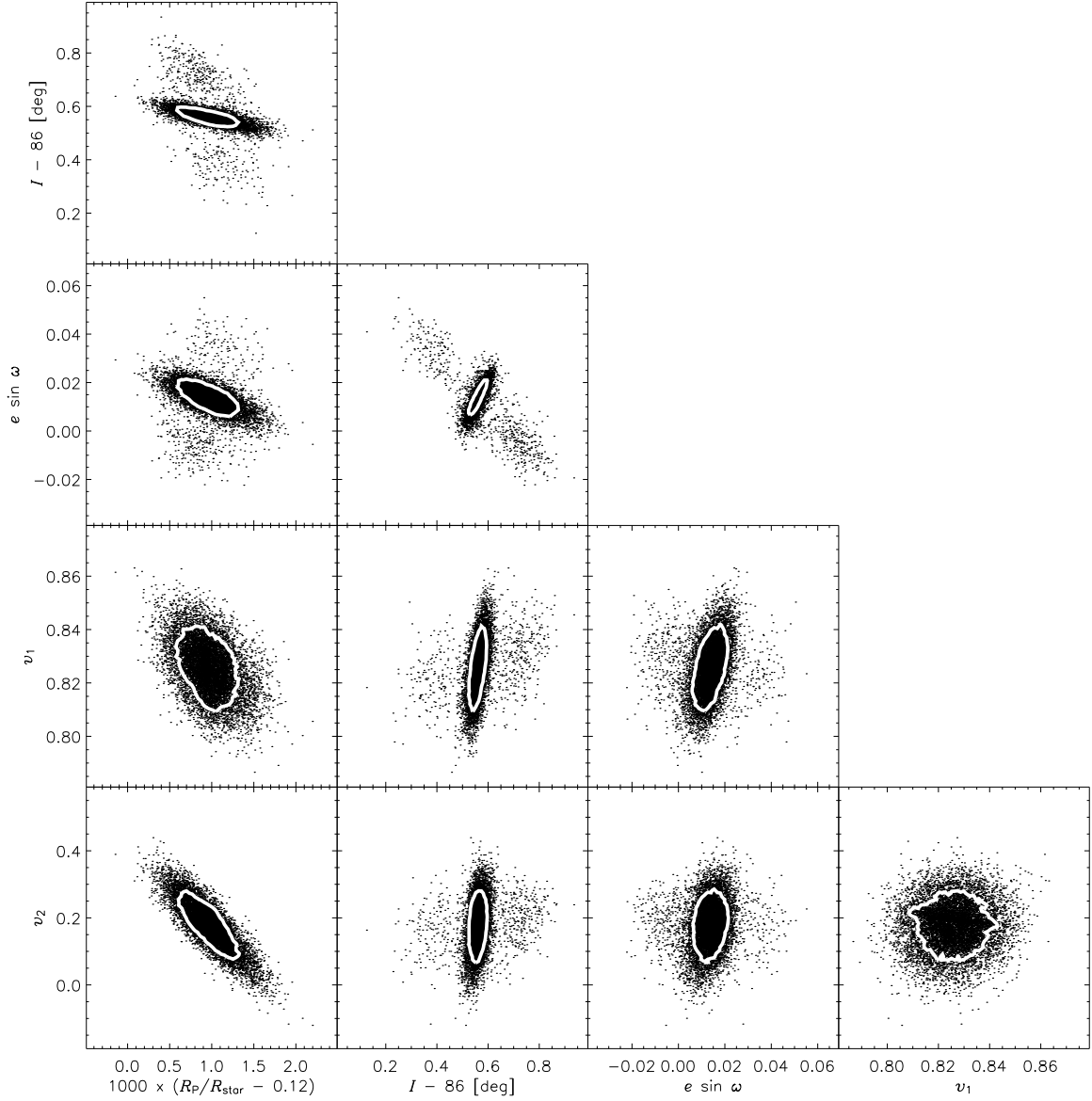}
\caption{ Joint probability distributions of some planetary, stellar,
and orbital parameters. The density of points is proportional to the
probability density. The white lines are isoprobability contours
enclosing 68\% of the points. Results are shown for the specific
choice $M_\star/M_\odot=1.06$, but none of the distributions plotted
varies significantly with stellar mass. The parameters shown here are
determined mainly by the photometry.
\label{fig:corr_lc}}
\end{figure}

\begin{figure}
\epsscale{1}
\plotone{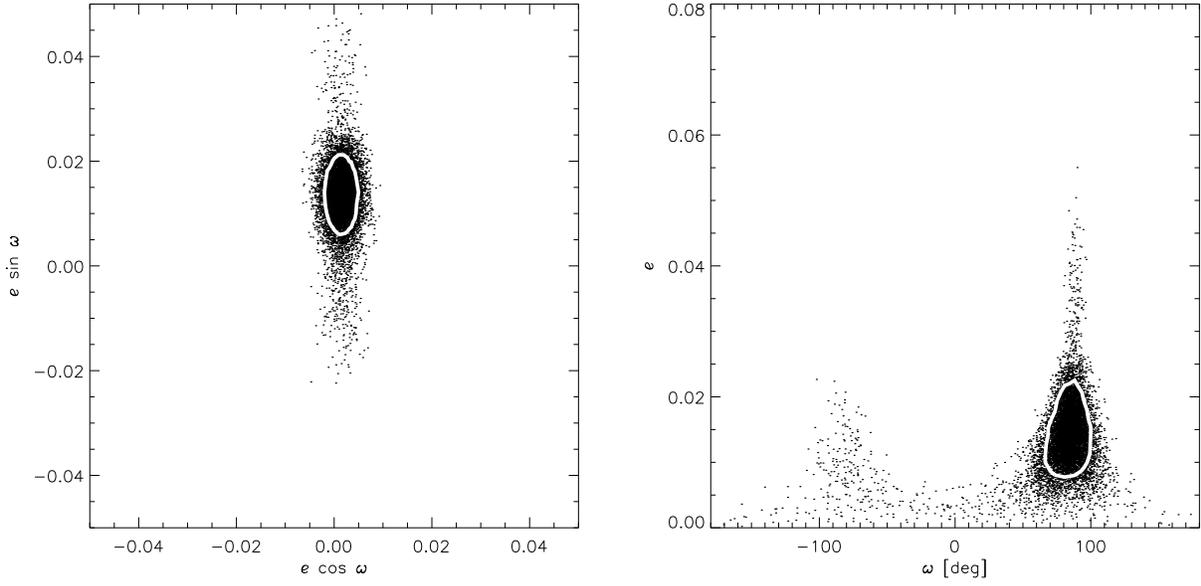}
\caption{ Same as Fig.~\ref{fig:corr_lc}, but for parameters relating
to the orbital eccentricity. The nonzero result for $e\sin\omega$ is
probably an artifact of the limb-darkening model (see
\S\ref{subsec:ecc}).
\label{fig:corr_e}}
\end{figure}

\begin{figure}
\epsscale{1}
\plotone{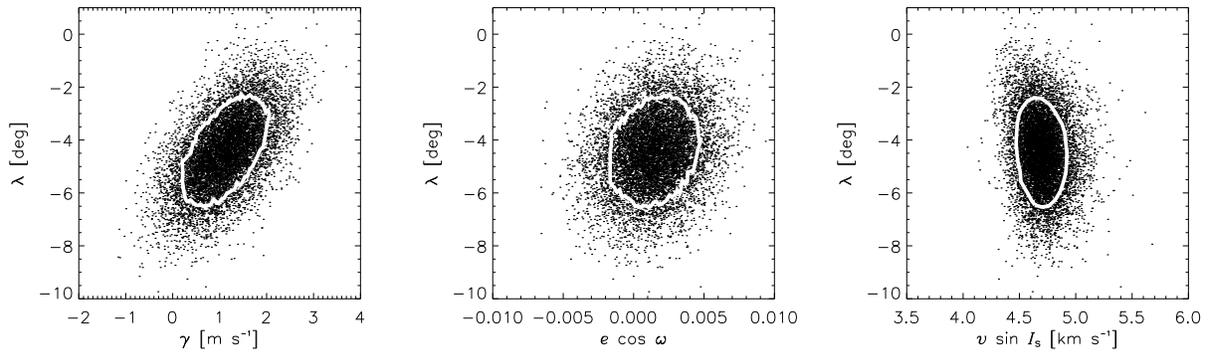}
\caption{ Same as Fig.~\ref{fig:corr_lc}, but for parameters relating
to the Rossiter-McLaughlin effect.
\label{fig:corr_ross}}
\end{figure}

Some of the parameters have correlated
uncertainties. Figures~\ref{fig:corr_lc}--\ref{fig:corr_ross} show
most of these correlations, for the particular choice of stellar mass
$M_{\star}/M_\odot=1.06$. Each panel shows the 12-dimensional
probability distribution function projected onto a two-dimensional
plane in parameter space. The isoprobability contours enclosing 68\%
of the results are plotted, to show the approximate ``1$\sigma$''
joint confidence regions for the two plotted parameters. In the
remainder of this section, we call attention to some of the principal
results.

\subsection{The Rossiter-McLaughlin effect}
\label{subsec:rm}

We find $v\sin I_\star= (4.70\pm 0.16)$~km~s$^{-1}$ for the
line-of-sight rotation speed of the star at the latitude crossed by
the planet.\footnote{It might appear that this result represents an
extrapolation beyond the grid of $\Delta v(f,v_p)$ that was the
basis of Eq.~(\ref{eq:jj}), but this is not so. For a transit at
stellar latitude $b$ and $\lambda\approx 0$, the maximum sub-planet
velocity is $v_p = v\sin I_\star \cos b$. In this case,
$b\approx 30\arcdeg$, and the maximum $v_p$ is 4.1~km~s$^{-1}$,
whereas our grid extended up to 4.5~km~s$^{-1}$.}  This result is in
agreement with (but is more precise than) the value $3.75\pm
1.25$~km~s$^{-1}$ found by Queloz et al.\ (2000). This number is
primarily determined by the amplitude of the radial velocity anomaly
and the depth of the transit, as interpreted through
Eq.~(\ref{eq:jj}). If the OTS formula is used instead of
Eq.~(\ref{eq:jj}), then we obtain $v\sin I_\star = 6.0$~km~s$^{-1}$;
but, as discussed in \S~\ref{subsec:rv}, the applicability of the OTS
formula to these data is questionable.

We find $\lambda = -4\fdg4 \pm 1\fdg4$, a small but significantly
nonzero angle. How robust is this result? This number is primarily
determined by the time interval between two events:\ the moment when
the radial velocity anomaly vanishes [$t(\Delta v = 0)$], and the
moment of greatest transit depth ($t_{\rm I}$). As long as $I\neq
90\arcdeg$ and $I_\star \neq 0\arcdeg$, then
\begin{equation}
\lambda = 0 \hspace{0.1in}
\longleftrightarrow \hspace{0.1in}
t(\Delta v = 0) = t_{\rm I}.
\end{equation}
Thus, our ability to test whether or not $\lambda$ is zero depends
chiefly on on our ability to assign consistent orbital phases to all
the radial velocities and photometry obtained during transits. It
depends secondarily on the radial velocities measured outside
transits, since those data determine the Keplerian orbit and allow the
radial velocity anomaly to be isolated. It does {\it not} depend
critically on our models for the limb darkening or the RM effect; in
any reasonable limb-darkening model, $t_{\rm I}$ occurs when the
projected planet--star separation is smallest, and in any reasonable
RM model, $t_{\Delta v=0}$ occurs when the planet crosses the
projected stellar rotation axis.

In the best-fitting model, shown in Fig.~\ref{fig:bestfit}, one can
see that the model radial velocity curve passes close to the origin,
but that there is a small offset between zero radial velocity
variation and zero photometric phase. The time interval $t(\Delta v =
0) - t_{\rm I}$ is 4.3 minutes. This timing offset is not caused by
the uncertainty in the orbital period $P$. The elapsed time between
the transit photometry and the transit subset of radial velocities was
approximately 3 months, or 25 orbits. The 83~ms uncertainty in $P$
causes an uncertainty of only $\sim$2 seconds in lining up the radial
velocity and the flux data, which is much smaller than 4.3
minutes. Even over the 5~yr time span of all the Laughlin et al.\
(2005) measurements, the maximum phase offset due to the uncertainty
in $P$ corresponds to a timing offset of only 43 seconds. If we fix
$\lambda=0\arcdeg$ and allow $P$ to vary to best fit the data, we find
$P=3.52472441$~days, which is 25$\sigma$ larger than the externally
measured period.

A more serious concern than the uncertainty in $P$ is the possibility
that the transit radial velocity measurements are all systematically
high because of stellar jitter. The time scales and corresponding
amplitudes of these random radial velocity excursions are poorly
known, but if the star happened to experience an excursion of
$\sim$5~m~s$^{-1}$ during the entire night of the transit of
2000~July~29, then our conclusion regarding a nonzero $\lambda$ would
be erroneous. The only way to settle this matter would be to obtain
additional radial velocity measurements during transits.

Two of the model parameters have variances that are notably correlated
with the variance in $\lambda$. One of them is $\gamma$, the velocity
zero point, which is easily understood since $t(\Delta v = 0)$ depends
on $\gamma$. The second correlated parameter is $e\cos\omega$, which
is relevant because it controls the timing offset between the moment
when the Keplerian radial velocity variation vanishes, and the moment
when the anomalous Doppler shift vanishes. These correlations are
shown in Fig.~\ref{fig:corr_ross}. Also plotted is the joint
distribution of $v\sin I_\star$ and $\lambda$, showing that the
variances in those parameters are not correlated.

\subsection{The masses and radii}

The best-fitting values for the mass and radius of the planet, and for
the radius of the star, are given in Table~1. The uncertainties in
these quantities are dominated by the uncertainty in the stellar
mass. These well-known degeneracies can be written $M_p \propto
M_\star^{2/3}$, $R_p \propto M_\star^{1/3}$, and $R_\star
\propto M_\star^{1/3}$. The ratio of radii does not depend on
$M_\star$ and is determined with high accuracy from the
photometry. Cody \& Sasselov (2002) used theoretical models of stellar
structure and evolution to show that $R_\star$ is a {\it decreasing}
function of $M_\star$, for fixed values of the star's observable
properties (luminosity, effective temperature, and metal
abundance). If one were willing to use the theoretical models to set
{\it a priori} constraints on the mass--radius relationship, then the
degeneracy would be partially broken.

\subsection{The limb darkening}

The two parameters that describe the limb darkening are $u_1$ and
$u_2$. The uncertainties in these parameters are highly correlated,
and one linear combination of these parameters is much more tightly
constrained than the orthogonal combination. The appropriate linear
combinations are
\begin{eqnarray}
v_1 & \equiv & u_2 + \frac{5}{3}u_1 = 0.83\pm 0.01, \\
v_2 & \equiv & u_2 - \frac{3}{5}u_1 = 0.18\pm 0.07.
\end{eqnarray}
The variances of these parameters are uncorrelated (see
Fig.~\ref{fig:corr_lc}). For this reason, Table~1 gives the results
for $v_1$ and $v_2$ rather than for $u_1$ and $u_2$. In contrast,
Brown et al.\ (2001) presented results for the parameters $u_1+u_2$
and $u_1-u_2$, which we find to be correlated. Some further remarks on
the effect of limb darkening are given below.

\subsection{The orbital eccentricity}
\label{subsec:ecc}

The 90\% confidence upper limit on the eccentricity is $e<0.023$. This
agrees with the upper limit that was achieved by Laughlin et al.\
(2005) using only the radial velocity data. But are the data
consistent with zero eccentricity? The answer to this question is
important because, as mentioned in \S~1, the observation of a nonzero
eccentricity might help to solve the mystery of why the planet's
density is much smaller than the density of Jupiter (Bodenheimer et
al.\ 2001). Formally, we find $e>0.0057$ with 90\% confidence (see
Fig.~\ref{fig:corr_e}), but this result is highly suspect. We believe
that the data are actually consistent with zero eccentricity, and that
the lower bound on $e$ is an artifact of our imperfect limb-darkening
model. Although this problem does not affect the RM results or any of
the discussion on which they are based (\S~\ref{sec:discussion}), we
digress here to explain the problem in more detail, and suggest how to
correct it in future analyses.

The natural parameters for describing departures from circularity are
$e\cos\omega$ and $e\sin\omega$, because the errors in these
parameters are uncorrelated, whereas $e$ and $\omega$ are correlated
(see Fig.~\ref{fig:corr_e}). In addition, the determination of $e$ is
biased because $e$ must be positive; random errors in measurements of
a circular orbit can only increase the apparent eccentricity. In
contrast, $e\cos\omega$ and $e\sin\omega$ can assume positive or
negative values. The quantity $e\cos\omega$ is independently
constrained by the measurement of the time of the secondary eclipse
(\S~\ref{subsec:secondary}), and is consistent with zero with a small
uncertainty. The quantity $e\sin\omega$ controls the speed of the
planet during transit and thus the duration of the transit (the time
between first and last contacts). However, the measured duration also
depends on several other parameters, namely $R_p$, $R_{\star}$, $I$,
$u_1$ and $u_2$. A bias in any of these parameters produces a
corresponding bias in $e\sin\omega$.\footnote{Another way to describe
the situation is that when $e$ is small and $\omega=\pm 90\arcdeg$,
the only observable consequence of $e\neq 0$ is a time-symmetric
distortion of the transit light curve: a dilation or compression that
is symmetric about the central transit time. Time-symmetric
distortions are also produced by varying the inclination and the
limb-darkening coefficients, which is why there is a degeneracy
between $e\sin\omega$ and those other parameters.} In particular, if
the quadratic limb-darkening law is not a perfect description of the
true limb darkening, then $e\sin\omega$ can be adjusted to compensate
for the imperfection.

If $e\cos\omega=0$ and $e\sin\omega$ is nonzero due to such a bias,
then relatively large values of $e$ are allowed when $\omega$ is
nearly $90\arcdeg$ or $-90\arcdeg$. This is precisely what is seen in
Fig.~\ref{fig:corr_e}. The high-$e$ solutions are also evident in
Fig.~\ref{fig:corr_lc} as the extended ``wings'' in the probability
distributions of the parameters that are correlated with
$e\sin\omega$. We confirmed that different choices for the
limb-darkening law result in significantly different values for
$e\sin\omega$. In short, the quoted uncertainties for the parameters
that describe the photometry are internal to our choice of
limb-darkening law. Because the present work is concerned mainly with
$\lambda$ and $v\sin I_\star$, which are not correlated with any of
those parameters, we have not attempted to correct this bias. One
could do so by adopting a more accurate or a more general model for
limb darkening, and by incorporating multi-color photometry rather
than the monochromatic photometry analyzed here. Alternatively, once
it becomes possible to measure the duration of secondary eclipse with
much greater precision, Eq.~(\ref{eq:secondary}) can be used to place
a direct constraint on $e\sin\omega$ (though there may be a related
bias in that measurement as well).

\section{Summary and Discussion}
\label{sec:discussion}

Through a joint analysis of all the best measurements of HD~209458
that are available, we have estimated the orbital, stellar, and
planetary parameters and their uncertainties. Our results agree with
previous determinations of these parameters by investigators who
analyzed subsets of these data, and in some cases we have modestly
decreased the uncertainties (cf.\ Brown et al.\ 2001, Deming et al.\
2005, Laughlin et al.\ 2005). The greatest improvement was achieved
for the parameters describing the Rossiter-McLaughlin effect, $v\sin
I_\star$ and $\lambda$, for which the uncertainties have been
decreased by a factor of 10 relative to the results of Queloz et al.\
(2000). This was accomplished by employing higher-precision radial
velocity measurements and by interpreting these measurements with a
new modeling technique.

The angle $\lambda$ is measured between the sky projections of two
vectors: ${\mathbf L}_p$, the orbital angular momentum of the
planet; and ${\mathbf L}_\star$, the rotational angular momentum of
the star. Thus, $\lambda$ is a lower bound on the three-dimensional
angle $\psi$ between these two vectors. The relation between $\lambda$
and $\psi$ depends on the inclinations $I$ and $I_\star$ of the
planetary orbit and stellar spin axis:
\begin{equation}
\cos\psi = \cos I_\star \cos I + \sin I_\star \sin I \cos\lambda.
\end{equation}
Since the orbit is nearly edge-on, this relation can be simplified to
$\cos\psi \approx \cos\lambda \sin I_\star$. We expect $\psi$ to be
not much larger than $\lambda$, unless the star's axis is pointed
towards the Earth (a coincidence that is {\it a priori} unlikely and
would also imply a stellar rotation speed $v$ that is considerably
faster than expected for a middle-aged G dwarf). In what follows, we
suppose that $\psi\lsim 0.1$~rad, and consider the implications.

This result is reminiscent of the planetary orbits in the solar
system. Relative to the net angular momentum vector of the solar
system, the rotation axis of the Sun is inclined by 6\arcdeg, and the
orbital angular momentum vectors of individual planets are tipped by
3--10\arcdeg. The planet orbiting HD~209458 is similar to solar system
planets in this respect, even though it is much closer to its parent
star than any of the planets in the solar system. Its orbital distance
is only about 9 stellar radii, as compared to 83 stellar radii for the
orbit of Mercury. In this sense, our result extends by a factor of 9
the range of orbital distances over which spin-orbit alignment has
been measured.

This proximity to its parent star raises the question of whether any
novel spin-orbit interactions can be observed in the HD~209458 system
that are not observed in the solar system. Miralda-Escud\'{e} (2002)
considered the interaction between a close-in giant planet and its
parent star's gravitational quadrupole field. If the planetary orbit
is inclined, the line of nodes of the orbit regresses. This effect is
potentially measurable as a slow secular variation in the duration of
transits. The precession frequency is
\begin{equation}
\dot{\Omega} = -\left(\frac{2\pi}{P}\right)
                \left(\frac{R_\star}{a}\right)^2
                \left(\frac{3J_2}{4}\right)
                \left(\frac{\sin 2\psi}{\sin\psi_p}\right),
\end{equation}
where $a$ is the semimajor axis of the planetary orbit (in this case,
$a/R_\star =8.65$), $J_2$ is the dimensionless quadrupole moment of
the star, and $\psi_p$ is the angle between ${\mathbf L}_{\rm
P}$ and ${\mathbf L}_p + {\mathbf L}_\star$. Assuming $J_2\sim
10^{-6}$ (a few times larger than the Sun's quadrupole moment), the
precession rate for HD~209458 is $\sim$$4\arcsec$~yr$^{-1}$,
corresponding to a precession period of $6\times 10^{7}$ orbital
periods. Successive transits should vary in duration by a fractional
amount $\sim$10$^{-8}$. This is a minuscule effect, but it may
nevertheless be detectable with high-precision photometry spanning
several years (see Miralda-Escud\'{e} 2002, \S~2.3).

Tidal interaction between the star and planet should also be
considered. The planet raises a tide on the star, and vice versa.
These tides dissipate energy, and the tidal bulges of one body exert
torques on the other body. Rasio et al.\ (1996) showed that the orbit
of 51~Peg~b (a typical close-in giant planet) is formally unstable to
tidal decay, but also that the timescale for orbital shrinkage is
longer than the main-sequence lifetime of the parent star. Likewise,
tidal dissipation acts to coplanarize the orbit and the stellar
equator (Greenberg 1974; Hut 1980), but we show presently that the
timescale for this process is longer than the age of the system. Using
a simplified model of tidal friction, Hut (1981) calculated the time
evolution of a general two-body orbit.  Specializing to the case
$e=0$, and considering first only the tide raised on the star by the
planet, the equation governing the evolution of $\psi$ can be written
(to first order in $\psi$):
\begin{equation}
\frac{1}{\psi} \frac{d\psi}{dt} =
-\frac{3k}{4\pi r_g^2 Q_\star}
\left( \frac{GM_\star}{R_\star^3} P_\star \right)
\left( \frac{M_p}{M_\star} \right)^2
\left( \frac{R_\star}{a} \right)^6
\left[ 1 - \frac{1}{2}\left(1 - \frac{L_\star}{L_p}\right)\frac{P}{P_\star} \right],
\end{equation}
where $P_\star$ is the rotation period of the star; $Q_\star$ is the
dimensionless ``quality factor'' of the tidal oscillations in the star
(and is inversely proportional to the dissipation rate); $k$ is the
apsidal motion constant ($k\approx$0.01 for a solar-type star); and
$r_g$ is the dimensionless radius of gyration
($r_g^2\approx$0.1). From this equation, the characteristic time scale
for significant change in $\psi$ is
\begin{equation}
\tau_\psi
\sim
\left( \frac{4\pi r_g^2 Q_\star}{3k} \right)
\left( \frac{R_\star^3}{GM_\star} \frac{1}{P_\star} \right)
\left( \frac{M_\star}{M_p} \right)^2
\left( \frac{a}{R_\star} \right)^6
\sim
5\times 10^{12} \hspace{0.03in} {\rm yr} \hspace{0.02in} \left( \frac{Q_\star}{10^6} \right),
\end{equation}
where we have evaluated the expression for the parameters of HD~209458
and a reasonable guess for $Q_\star$ (Terquem et al.\ 1998). Since the
age of the star is only about $5\times 10^9$~yr (Cody \& Sasselov
2002), any inclination damping should be negligible. The inclination
we have observed is therefore likely to be a relic of the planet
formation epoch. Alternatively, our measurement of a nonzero
inclination could be considered as a weak upper bound on tidal
dissipation, $Q_\star < 10^{10}$.

The star also raises tides on the planet. Through this mechanism, the
planet's rotation period is synchronized with the orbital period and
the orbit is circularized on relatively short timescales ($\tau_{\rm
synch} \sim 10^6$~yr and $\tau_{\rm circ} \sim 10^9$~yr; Rasio et al.\
1996). This is why the perturbing effect of a third body is required
to maintain a nonzero eccentricity, in the scenario proposed by
Bodenheimer et al.\ (2001). However, if the orbit is inclined relative
to the stellar quadrupole, then the planet experiences a time-variable
tidal distortion even after synchronization and circularization are
achieved. The planet makes a vertical oscillation in the quadrupolar
field of the star once per orbit. To investigate whether this
oscillation produces a significant amount of heat, we followed the
same procedure that Peale \& Cassen (1978) used to calculate tidal
heating within the Moon, and that Wisdom (2004) recently applied to
the case of Enceladus. We identified the leading time-variable term in
the tidal potential and calculated the resulting height of the tide,
approximating the planet as an incompressible fluid. For the case of
HD~209458b, the time-averaged heating rate is
\begin{equation}
\frac{dE}{dt} =
2\times 10^{11} \hspace{0.03in} {\rm erg~s}^{-1} \hspace{0.03in}
\left( \frac{Q_p}{10^6} \right)^{-1}
\left( \frac{J_2}{10^{-6}} \right)^2
\left( \frac{\sin\psi}{0.1} \right)^2 \nonumber =
6\times 10^{-16} \hspace{0.05in}
\left(
\frac{GM_p^2 / R_p}{10^9 \hspace{0.02in} {\rm yr}}
\right),
\end{equation}
to leading order in $J_2 \sin\psi$. This heating rate is utterly
negligible compared to other sources of heat such as gravitational
contraction and stellar insolation. A potentially more powerful source
of heat is the tide that would be produced by a nonzero planetary
obliquity. For a close-in giant planet with $\psi\neq 0$, it may be
possible for the obliquity (the angle between the planetary spin axis
and the orbit normal) to avoid being driven to zero during the
synchronization process, a theoretical possibility that has been
explored by Winn \& Holman (2005).

Finally, we turn to the implications of our measurement of the stellar
spin, $v \sin I_\star = (4.70\pm 0.16)$~km~s$^{-1}$. Because $\lambda$
is nearly zero, the result can be interpreted as the line-of-sight
rotation speed of the star along the stellar latitude traversed by the
transiting planet. The stellar spin can also be estimated in the
traditional manner, by interpreting the observed broadening of
photospheric absorption lines. Fischer \& Valenti (2005) did so for
HD~209458 using Keck/HIRES spectra, finding $v\sin I_\star = (4.5 \pm
0.3)$~km~s$^{-1}$. Likewise, Mazeh et al.\ (2000) reported two
estimates of $v\sin I_\star$, $(4.4\pm 1.0)$ and $(4.1\pm
0.6)$~km~s$^{-1}$, based on spectra taken with two different
instruments, and Shkolnik et al.\ (2005) found $v\sin I_\star=(4.2\pm
0.5)$~km~s$^{-1}$. All these values are in agreement within the quoted
uncertainties.

The agreement is potentially interesting for at least three
reasons. First, it is a consistency check on our method for
interpreting the Rossiter-McLaughlin effect
(\S~\ref{subsec:rv}). Second, assuming that our method is correct, it
provides a rare example apart from the Sun for which the traditional
interpretation of spectral line broadening can be checked. There are
physical effects besides rotation that contribute to line broadening,
such as macroturbulence and microturbulence. In general, assumptions
must be made about the magnitude of these other effects in deriving
$v\sin I_\star$ from the net observed line broadening. Our result
shows that for the case of HD~209458, these assumptions are apparently
justified. Finally, if our method and the traditional interpretation
of line broadening are assumed to be correct, then the agreement
between the two estimates of $v\sin I_\star$ places an upper bound on
any differential rotation of the star. The line-broadening measurement
is a disk-averaged quantity, whereas the transit measurement refers
specifically to a latitude of $30\arcdeg$. Together, the two
measurements imply $(v_{\rm avg} - v_{\rm 30\arcdeg}) \sin I_\star =
-0.20 \pm 0.34$~km~s$^{-1}$, corresponding to $(-4.4 \pm 7.6)$\% of
the average rotation speed. Unfortunately, we lack the precision to
detect the degree of differential rotation expected of a solar-type
star. On the Sun, differential rotation between the equator and
latitude $30\arcdeg$ is only about 0.1~km~s$^{-1}$, or 5\% of the
equatorial speed (Beck 2000), and of course the difference between the
disk-averaged rotation and the rotation at latitude 30\arcdeg\ is even
smaller.

In closing, we wish to point out that although HD~209458 is presently
the only extrasolar planetary system for which the Rossiter-McLaughlin
effect has been detected, it is possible that a large sample of
suitable systems will soon be available. A second example of a
transiting planet with a bright parent star was recently discovered:
TrES-1, whose parent star is a 12th magnitude K~dwarf (Alonso et al.\
2004). It would be interesting to perform a similar analysis of this
system, given the different stellar type and planetary
characteristics. Numerous wide-field surveys for transiting planets
are underway, which we hope will provide a bounty of additional
targets in the near future.

\acknowledgments We acknowledge helpful discussions with T.\ Brown,
E.\ Chiang, S.\ Gaudi, D.\ Lin, and G.\ Torres. We are grateful to J.\
Wisdom for advice on calculating the rate of tidal heating. Work by
J.N.W.\ was supported by NASA through Hubble Fellowship grant
HST-HF-01180.02-A, awarded by the Space Telescope Science Institute,
which is operated by the Association of Universities for Research in
Astronomy, Inc., for NASA, under contract NAS~5-26555. The visit of
E.L.T.\ to the University of Tokyo was supported by an invitation
fellowship program for research in Japan from the Japan Society for
Promotion of Science (JSPS). This work was partly supported by a
Grant-in-Aid for Scientific Research from JSPS grants 14102004 and
16340053, and by NASA grant NAG5-13148.

\begin{deluxetable}{cccccc}

\tabletypesize{\scriptsize}
\tablecaption{Orbital, stellar, and planetary properties of HD~209458\label{tbl:params}}
\tablewidth{0pt}

\tablehead{
\colhead{} &
\colhead{Best Fit} &
\colhead{Uncertainty} &
\colhead{Lower 90\%} &
\colhead{Upper 90\%} &
\colhead{}\\
\colhead{Parameter} &
\colhead{(mean)} &
\colhead{($\sigma$)} &
\colhead{Confidence Limit} &
\colhead{Confidence Limit} &
\colhead{Notes}
}
\startdata
              $M_p$ ($M_{\rm Jup}$) & $    0.657$   & $    0.006$   & $    0.647$   & $    0.668$    & 1 \\
              $M_p$ ($M_{\rm Jup}$) & $    0.657$   &     \nodata   & $    0.594$   & $    0.721$    & 2 \\
              $R_\star$ ($R_\odot$) & $    1.148$   & $    0.002$   & $    1.143$   & $    1.152$    & 1,3 \\
              $R_\star$ ($R_\odot$) & $    1.15$    &     \nodata   & $    1.09$    & $    1.20$     & 2,3 \\
              $R_p$ ($R_{\rm Jup}$) & $    1.355$   & $    0.002$   & $    1.350$   & $    1.358$    & 1,3 \\
              $R_p$ ($R_{\rm Jup}$) & $    1.35$    &     \nodata   & $    1.29$    & $    1.41$     & 2,3 \\
                      $R_p/R_\star$ & $    0.12096$ & $    0.00025$ & $    0.12056$ & $    0.12141$  &  \\
                      $e\cos\omega$ & $    0.0014$  & $    0.0022$  & $   -0.0021$  & $    0.0049$   &  \\
                      $e\sin\omega$ & $    0.0141$  & $    0.0055$  & $    0.0037$  & $    0.0232$   & 3 \\
                                $e$ & $    0.0147$  & $    0.0053$  & $    0.0057$  & $    0.0234$   & 3 \\
                     $\omega$ (deg) & $   84$       & $   11$       & $   56$       & $   99$        & 3 \\
              $\gamma$ (m~s$^{-1}$) & $    1.11$    & $    0.63$    & $    0.08$    & $    2.12$     &  \\
           $\Delta t_{\rm I}$ (sec) & $   -5.7$     & $    2.0$     & $   -9.0$     & $   -2.6$      &  \\
                          $I$ (deg) & $   86.55$    & $    0.03$    & $   86.49$    & $   86.61$     & 3 \\
      $v\sin I_\star$ (km~s$^{-1}$) & $    4.70$    & $    0.16$    & $    4.44$    & $    4.97$     &  \\
                    $\lambda$ (deg) & $   -4.4$     & $    1.4$     & $   -6.8$     & $   -2.1$      &  \\
  $v_1 \equiv u_2 + \frac{5}{3}u_1$ & $    0.825$   & $    0.010$   & $    0.808$   & $    0.842$    & 3 \\
  $v_2 \equiv u_2 - \frac{3}{5}u_1$ & $    0.181$   & $    0.074$   & $    0.058$   & $    0.289$    & 3
\enddata

\tablenotetext{1}{Based on the assumption $M_\star/M_\odot = 1.06$.}

\tablenotetext{2}{Incorporates the uncertainty in the stellar mass.
The lower confidence limit is for $M_\star/M_\odot=0.93$, and the
upper confidence limit is for $M_\star/M_\odot=1.19$.}

\tablenotetext{3}{Depends on our particular choice of limb-darkening
law. In reality, $e$ is probably consistent with zero (see \S~4.4).}

\end{deluxetable}

\end{document}